\documentclass[10pt]{iopart}

\usepackage{iopams}
\usepackage{bm}
\usepackage{graphicx}
\usepackage{comment}
\usepackage{epsfig}                  

\renewcommand{\>}{\right \rangle}

\newcommand{\ket}[1]{\left |#1\>}

\newcommand{\be}{\begin{equation}}
\newcommand{\ee}{\end{equation}}
\newcommand{\bea}{\begin{eqnarray}}
\newcommand{\eea}{\end{eqnarray}}

\newcommand{\aver}[1]{\langle #1 \rangle}

\begin{document}
\title[Non-markovian dynamics in atom-laser outcoupling from a double-well BEC]{Non-Markovian 
dynamics in atom-laser outcoupling from a double-well Bose-Einstein condensate}

\author{C. Lazarou$^{1}$, G. M. Nikolopoulos$^{2}$, P. Lambropoulos$^{2,3}$}

\address{$^1$ Department of Physics and Astronomy, University of Sussex, 
Brighton BN1 9QH, United Kingdom}

\address{$^2$ Institute of Electronic Structure and Laser, FORTH, P. O. Box 1527, Heraklion 711 10, Crete, Greece}

\address{$^3$ Department of Physics, University of Crete, P. O. Box 2208, Heraklion 71003, Crete, Greece}


\begin{abstract}
We investigate the dynamics of a continuous atom laser based on the merging 
of independently formed atomic condensates. In a first attempt to understand 
the dynamics of the system, we consider two independent elongated 
Bose-Einstein condensates which approach each other and focus on intermediate 
inter-trap distances so that a two-mode model is well justified. 
In the framework of a mean-field theory, we discuss the quasi steady-state 
population of the traps as well as the energy distribution of the outcoupled 
atoms. 
\end{abstract}



\maketitle

\section{Introduction}
The theory of quantum dissipative systems, and in particular those in which 
structured reservoirs are involved, represents a currently active field of 
research in a rather broad and varied context \cite{bookBP}.  
The fundamental challenge in 
problems involving this type of reservoir stems from the inapplicability of 
the Born and Markov approximations, normally valid for a reservoir with a 
smooth density of states, coupled weakly to a quantum system with few degrees 
of freedom.  As a result, the so-called pole approximation which leads to the 
elimination of the reservoir degrees of freedom can not be adopted. On the 
other hand, there are no established approaches of general use capable of 
addressing all problems of this type. The situation is further complicated by 
the large variety of and seemingly unrelated physical contexts in which such 
problems may appear.  The mathematical structure of the density of states of 
the reservoir and the strength of the coupling to the quantum system are the 
determining factors and source of difficulty. Actually, their combination 
determines the so-called spectral response whose form in each case sets the 
rules as to which approach may be helpful.

The theoretical description of what is known as atom laser represents a most 
recent example of this class of problems. In fact, the features that make 
this problem non-Markovian resemble those of an excited atom inside a material 
with a photonic bandgap \cite{PBGreview}; 
even though the underlying physics is rather different.  
In analogy to optical lasers, 
atom lasers can be obtained by outcoupling atoms from a trapped 
Bose-Einstein condensate (BEC) to free space. 
The most crucial prerequisite for the realization of a continuous 
atom laser is the pumping mechanism replenishing the trapped condensate 
as atoms are outcoupled from it. 
A number of sophisticated techniques have been proposed 
to this end \cite{pumping}, and most of them rely on optical pumping 
between various internal and external atomic states. 
However, none of these techniques has been able to overcome 
intrinsic losses in the system and achieve laser action.
Up to date, perhaps the most promising scheme towards the realization 
of a continuous source of condensed atoms  
was demonstrated by Chikkatur {\em et al.} \cite{MITscience02}, 
and relies on the use of optical tweezers for the transport 
and the merging of independently produced BECs. 

A large number of theoretical models have been used in studies 
of atom lasers which rely mainly on Born-Markov master equations 
\cite{wiseman,mey} 
and Gross-Pitaevskii theory \cite{bal96,adams,jap00,savage2001,Graham, Edwards}. In either case, the models hold only under 
certain operating conditions. For instance, it is well known that for experimentally 
achievable parameters atom lasers may exhibit non-Markovian dynamics 
\cite{moypra299,rmhc05}, which cannot be described in the 
framework of Born-Markov approximations. Besides, inclusion of 
non-Markovian effects in the Gross-Pitaevskii theory is a rather 
difficult task. Hence, investigations of atom-laser dynamics beyond 
Born and Markov approximations have been mainly performed in the framework of 
a particularly simple model involving a 
single-mode condensate (trap mode) coherently coupled to a continuum 
of free-space modes 
\cite{moypra299,hopepra97,moypra97,hopepra00,jackpra99,bre99,jefpra00,NLP03}.

Motivated by the experiments of Chikkatur {\em et al.} \cite{MITscience02}, 
in the present work we extend these studies to a two-mode scenario. 
In particular, we consider two independent BECs consisting  
of a large number of bosonic atoms cooled into the lowest eigenmode of 
the corresponding trap. To account for the merging process, the two 
traps are brought together, while atoms are coherently outcoupled 
from one of the BECs only. 
We focus on an intermediate stage of the merging process 
where the separation of the two traps is so large that, on the one hand 
a two-mode model can be adopted, while on the other hand  
a coherent Josephson coupling is established between the two BECs. 
Our purpose is to investigate how the presence of the second 
trap mode affects the dynamics of the atom laser and in particular the 
distribution of the outcoupled atoms.  

\section{The system}
\label{SecII}
Our system (see figure \ref{system.fig}) consists of two independently 
prepared elongated BECs 
(A and B) and let $N$ be the total number of atoms in the system. 
The two traps are initially far apart and each BEC experiences 
only its local potential $V_{\rm A(B)}^{\rm (L)}({\bf r})$, while   
only the lowest level of each trap (condensate mode) is populated.
To allow for the merging of the two BECs, the two traps are brought 
together along one of the tightly confining radial directions.  
Simultaneously, atoms are outcoupled coherently from BEC A, 
by applying external electromagnetic fields. In this section  
we describe in detail the modeling of the system used throughout this work.

\subsection{Double-well potential}
\begin{figure}
  \begin{center}
    \leavevmode
    \epsfxsize8.5cm
    \epsfbox{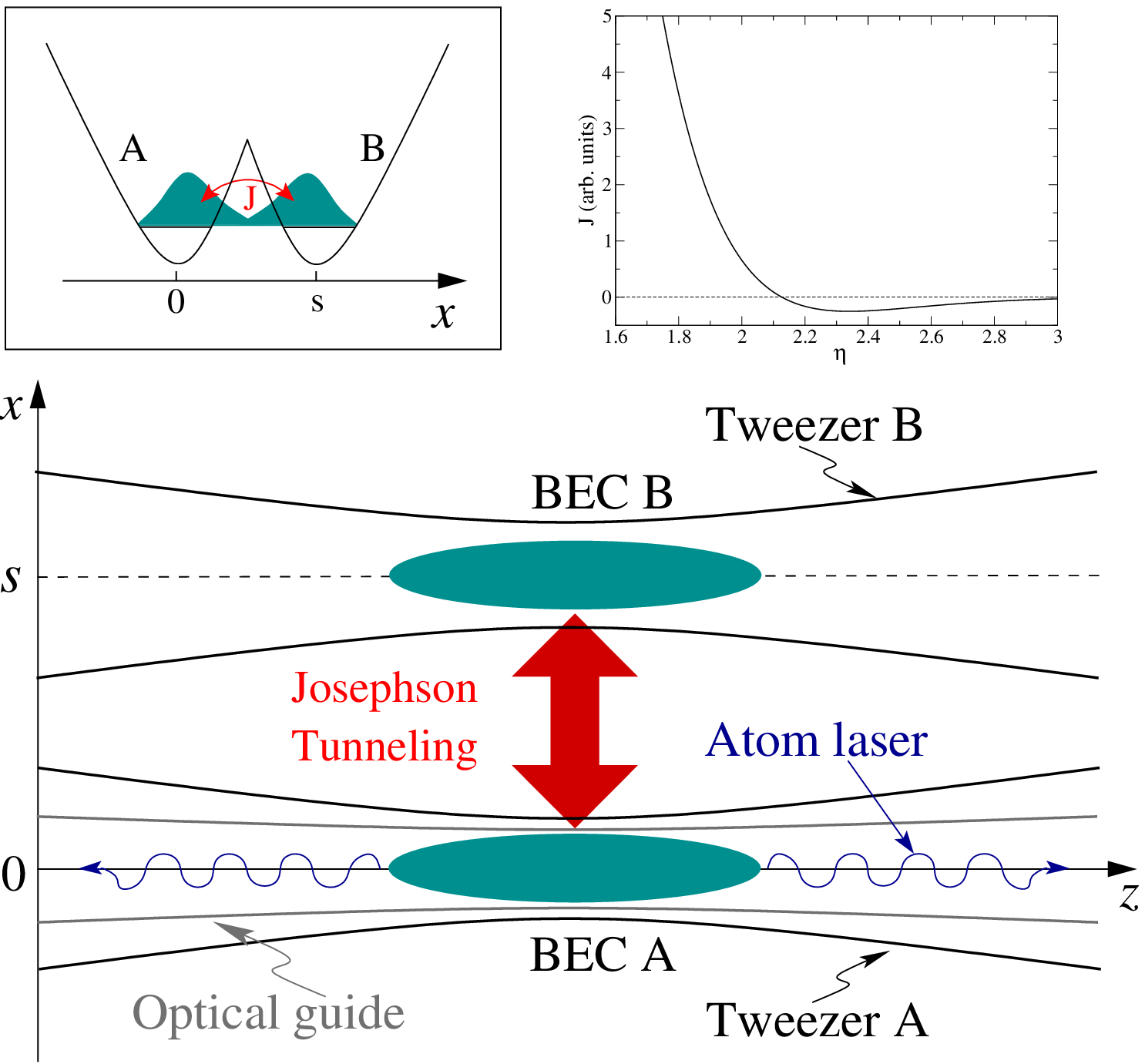}
  \end{center}
  \caption{Schematic representation of the system under consideration. 
    Left inset: The double-well potential experienced by the trapped atoms along the 
    $x$-direction. Right inset: The Josephson coupling as a function of the 
    dimensionless intertrap distance $\eta$.}
\label{system.fig}
\end{figure}

Transport of BECs can be realized using optical tweezers which are 
produced by focused laser beams and offer limited trap volume and depth.  
Hence, during the merging process the two BECs can be brought as close 
as the traps' beam waist, before they start affecting each other. 
To be consistent with the experimental setup for BEC merging 
\cite{MITscience02} 
as well as related theoretical work \cite{YiDuanpra05,MeSaLepra06}, 
we will assume two nearly identical 
axially symmetric harmonic traps with confining 
frequencies $\omega_z$ and $\omega_x=\omega_y=\omega_\perp$. 
Trap B is moving towards trap A along the radial direction $x$ 
(see figure \ref{system.fig}), and the global potential 
experienced by the trapped atoms can be modeled by a time-dependent 
double-well potential of the form, 
\bea
V_{\rm t}^{\rm (G)}({\bf r},t)
&=&\frac{1}{2}m\omega_x^2\left [\bigg |x-\frac{s(t)}{2}\bigg |-
\frac{s(t)}{2}\right ]^2+
\frac{1}{2}m\omega_y^2 y^2+
\frac{1}{2}m\omega_z^2 z^2,
\label{dw_pot}
\eea
where $m$ is the atomic mass. According to (\ref{dw_pot}),  
the harmonic potential remains unaffected in both 
$y$ and $z$ directions, while along the merging direction we have 
a double well potential which at any time $t$, exhibits two minima at 
$x=0$ and $x=s(t)$ (see left inset of figure \ref{system.fig}). 
The distance between the two dips decreases with time 
and at the end of the merging (i.e., at $t=t_{\rm m}$) 
we have complete overlap. 

Besides the merging time-scale $t_{\rm m}$, the details of the motion 
of trap B are not of great importance 
\cite{MITscience02,YiDuanpra05,MeSaLepra06}. 
The crucial point is that the BEC merging must be 
adiabatic so that any kind of excitations in the system are suppressed. 
To this end, first of all the transport of the BECs must take place on a 
time scale much larger than the characteristic time scale of excitations 
along the merging direction 
i.e., $t_{\rm m}\gg \omega_x^{-1}$ \cite{YiDuanpra05,MeSaLepra06}. 
Although this condition  
can be easily satisfied in a typical merging experiment \cite{MITscience02}, 
it does not ensure adiabaticity with respect to the time-scale of 
interatomic interactions \cite{YiDuanpra05}. Nevertheless, 
as long as $N U_{\rm tt} t_{\rm m}\gg 2\hbar V_{\rm mode}$, 
where $U_{\rm tt}$ is the strength of the interactions and 
$V_{\rm mode}$ is the effective mode volume for each trap, 
it has been shown that only 
low-lying eigenstates of the Hamiltonian may be populated 
during the merging \cite{YiDuanpra05}. 
As a result, at the end of the process one obtains a large single-mode 
BEC fraction with a unique relative phase. 

The local potentials centered at $x=0$ 
and $x=s(t)$, are readily obtained from equation (\ref{dw_pot}) 
\bea
&&V_{\rm A}^{\rm (L)}({\bf r})=
\frac{1}{2}m\left (\omega_x^2 x^2+\omega_y^2 y^2+\omega_z^2 z^2\right )
,\nonumber\\
&&V_{\rm B}^{\rm (L)}({\bf r},t)=
\frac{1}{2}m\left \{\omega_x^2 [x-s(t)]^2+\omega_y^2 y^2+
\omega_z^2 z^2\right \}.
\eea
In the case of an ideal bosonic gas, the wavefunction of the 
local ground states $\ket{{\rm A}}$ and $\ket{{\rm B}}$, corresponding to 
$V_{\rm A}^{\rm (L)}({\bf r})$ and $V_{\rm B}^{\rm (L)}({\bf r})$ 
respectively, have the well known  Gaussian-like profile i.e., 
\bea
\label{phi_a}
&&\varphi_{\rm A}({\bf r})\equiv\langle {\rm A}\ket{{\bf r}}=
\frac{1}{\pi^{3/4}l_z^{1/2}l_\perp}\exp\left [ 
-\frac{1}{2}\left (\frac{x^2+y^2}{l_\perp^2}+
\frac{z^2}{l_z^2}\right )\right ],
\nonumber\\
&&\varphi_{\rm B}({\bf r},t)\equiv\langle {\rm B}\ket{{\bf r}}=
\frac{1}{\pi^{3/4}l_z^{1/2}l_\perp}\exp\left \{ 
-\frac{1}{2}\left [\frac{[x-s(t)]^2+y^2}{l_\perp^2}+
\frac{z^2}{l_z^2}\right ]\right \}.
\label{phi_b}
\eea
The characteristic harmonic oscillator length $l_{\zeta}$ is defined as 
$l_{\zeta}\equiv\sqrt{\hbar/m\omega_\zeta}$ for $\zeta\in\{x,y,z\}$, 
while $l_\perp=l_x=l_y$. As we will see later on, these Gaussian-like 
profiles enable us to obtain analytic expressions for most of 
the parameters characterizing the dynamics of the system. From now 
on, for the sake of brevity we simply write $\varphi_{\rm B}({\bf r})$ 
instead of $\varphi_{\rm B}({\bf r}, t)$.

\subsection{Trapped atoms}
The many-body Hamiltonian describing the dynamics of the trapped atoms 
is given by \cite{jap00,book,TwoModeValid}
\bea
\fl\hat{\cal H}_{\rm t} &=& \int d{\bf r}\hat{\Psi}_{\rm t}^\dag({\bf r})
\left [-\frac{\hbar^2}{2m}\nabla^2+
V_{\rm t}^{\rm (G)}({\bf r},t)\right ]\hat{\Psi}_{\rm t}({\bf r})
+\frac{U_{\rm tt}}{2}\int d{\bf r}  
\hat{\Psi}_{\rm t}^\dag({\bf r}) \hat{\Psi}_{\rm t}^\dag({\bf r})
\hat{\Psi}_{\rm t}({\bf r}) \hat{\Psi}_{\rm t}({\bf r}),
\label{Ht0}
\eea
where $\hat{\Psi}_{\rm t}({\bf r})$ is the annihilation field operator for the 
trapped atoms with 
\be
\left [ \hat{\Psi}_{\rm t}({\bf r}), \hat{\Psi}_{\rm t}^\dag({\bf r}^\prime)
\right ] =\delta({\bf r}-{\bf r}^\prime),\quad 
\left [ \hat{\Psi}_{\rm t}({\bf r}), \hat{\Psi}_{\rm t}({\bf r}^\prime)\right ]=0.
\ee 
The quantity $U_{\rm tt}=4\pi\hbar^2a_{\rm tt}/m$ measures the strength 
of the interparticle interaction between trapped atoms, while $a_{\rm tt}$ 
is the corresponding $s$-wave scattering length.

At $t=0$, each BEC experiences only its local potential 
$V_{\rm A(B)}^{\rm (L)}$ as the two traps are well separated, 
and only the lowest level of each trap (condensate mode) is populated. 
We may expand therefore the field operator at $t=0$ as 
$\hat{\Psi}_{\rm t}({\bf r},0)=\varphi_{\rm A}({\bf r})\hat{a}(0)+
\varphi_{\rm B}({\bf r})\hat{b}(0)$, where $\varphi_{\rm A}({\bf r})$ and 
$\varphi_{\rm B}({\bf r})$ are the ground-state wavefunctions 
for the traps A and B, respectively \cite{jap00,book,TwoModeValid}. 
The corresponding bosonic 
annihilation operators are denoted by $\hat{a}$ and $\hat{b}$ and 
satisfy the standard commutation relations i.e., 
$\large [ \hat{d}_i, \hat{d}_j^\dag\large ]=\delta_{i,j}$, 
for $\hat{d}_1=\hat{a}$ and $\hat{d}_2=\hat{b}$.

To be consistent with the experiment \cite{MITscience02}, we will assume that 
during the merging the atomic density profiles 
follow adiabatically the movement of the traps. The condensate 
wavefunctions start overlapping in space as the two traps approach 
each other and a Josephson-type tunneling is established between the 
two BECs. If the position uncertainty in the ground state of the 
traps is much smaller than the separation of the minima of the global 
potential $V_{\rm t}^{\rm (G)}({\bf r},t)$ i.e., if 
\be
l_x \ll \sqrt{2}s,
\label{cond1}
\ee 
the overlap (and thus the Josephson coupling) is small enough so that only the ground 
states of the traps are relevant. In first-order perturbation theory, 
the corresponding local ground-state wavefunctions 
$\varphi_{\rm A(B)}({\bf r})$ are orthogonal and describe faithfully 
BEC A and B, at any time $0<t\ll t_{\rm m}$ \cite{TwoModeValid}. 
Hence, under such conditions we may still expand the field operator at times 
$0<t\ll t_{\rm m}$ as \cite{jap00,book,TwoModeValid}
\be
\hat{\Psi}_{\rm t}({\bf r},t)=\varphi_{\rm A}({\bf r})\hat{a}(t)+
\varphi_{\rm B}({\bf r})\hat{b}(t),
\label{Psi_t}
\ee
and the many-body Hamiltonian (\ref{Ht0}) 
reduces to the standard two-mode model 
\bea
\fl \hat{\cal H}_{\rm t}(t) &=& \hbar\omega_{\rm A}\hat{a}^\dag\hat{a}+
\hbar\omega_{\rm B} \hat{b}^\dag\hat{b}
+\hbar J\hat{a}^\dag\hat{b}+\hbar J^*\hat{b}^\dag\hat{a}
+\hbar \kappa_{\rm A} \hat{a}^\dag\hat{a}^\dag\hat{a}\hat{a}+
\hbar \kappa_{\rm B} \hat{b}^\dag\hat{b}^\dag\hat{b}\hat{b},
\label{Ht2}
\eea
where the coefficients are given by  
\bea
\fl \omega_j(t)&=&\frac{1}\hbar \int d{\bf r}\varphi_j^{*}({\bf r})
L_{\rm t}\varphi_j({\bf r}),
\quad
\kappa_{j}=\frac{U_{\rm tt}}{2\hbar}
\int d{\bf r}|\varphi_{j}({\bf r})|^4,\quad  
J(t)=\frac{1}\hbar \int d{\bf r}\varphi_{\rm A}^{*}({\bf r})
L_{\rm t}\varphi_{\rm B}({\bf r}),
\label{integrals:eq}
\eea 
with 
\bea
\label{Lt_glob}
L_{\rm t}(t) &=&-\frac{\hbar^2}{2m}\nabla^2 
+V_{\rm t}^{\rm (G)}({\bf r},t).
\eea 
In deriving equation (\ref{Ht2}) we have neglected 
higher-order cross-interaction terms involving integrands of the form 
$|\varphi_i({\bf r})|^2|\varphi_j({\bf r})|^2$, 
$[\varphi_i^{*}({\bf r})\varphi_j({\bf r})]^2$,
and $|\varphi_j({\bf r})|^2\varphi_i({\bf r})\varphi_j^*({\bf r})$. 

Condition (\ref{cond1}) itself does not justify completely the 
use of the two-mode model. In addition we have to guarantee that 
the effect of interatomic interactions on the ground-state properties 
of the two wells is small i.e., that  
$\hbar(\omega_x\omega_y\omega_z)^{1/3}\gg N\kappa_j$ \cite{TwoModeValid}.
For such weakly-interacting bosonic gases, the ground-state wavefunctions 
are well approximated by equations (\ref{phi_a}). The Gaussian profile of the 
wavefunctions enables us 
to evaluate analytically all of the integrals (\ref{integrals:eq}) 
obtaining for the coefficients entering the Hamiltonian 
$\hat{\cal H}_{\rm t}$ 
\be
\omega_j=\omega_{\rm o},\quad \kappa_j=\kappa,\quad J=J^*, 
\ee
where 
\bea
&&\omega_{\rm o}(t)=\omega_z\left [\frac{1}2+\frac{1}{\lambda} 
+\frac{\eta^2}{\lambda}\textrm{Erfc}(\eta)
-\frac{\eta}{\lambda\sqrt{\pi}}e^{-\eta^2}\right ],
\label{et_eq}
\\
&&J(t)=\omega_z\left (\frac{1}2+\frac{1}{\lambda} 
-\frac{\eta}{\lambda\sqrt{\pi}}\right )e^{-\eta^2},
\label{Jt_eq}
\\
&&\kappa=\frac{\hbar a_{tt}}{\lambda m\sqrt{2\pi}l_z^3},
\label{kap_eq}
\eea
while $\textrm{Erfc}(\eta)$ is the complementary error function  
and we have introduced the dimensionless quantities 
$\eta(t)=s(t)/(2l_x)$ and $\lambda=\omega_z/\omega_x$. 
Condition $\hbar(\omega_x\omega_y\omega_z)^{1/3}\gg N\kappa_j$ thus yields 
the following upper bound on the total number of atoms we may consider 
\be
N\ll\lambda^{1/3}\sqrt{2\pi}\frac{l_z}{a_{\rm tt}}.
\label{Nmax}
\ee

For the reasons we discussed earlier in this section, throughout our 
simulations we focus on inter-trap distances $\eta\geq 1.5$. 
According to (\ref{Jt_eq}),  $J(t)$ is practically zero for large $\eta$ and 
increases in absolute value, as we bring the traps closer i.e., 
for decreasing $\eta$ 
(see right inset of figure \ref{system.fig}).
On the contrary, the  ground-state frequency $\omega_{\rm o}(t)$  
does not vary appreciably in the same regime of inter-trap distances and 
thus throughout our simulations we may safely assume that 
$\omega_{\rm o}$ remains practically constant i.e., 
\be
\omega_{\rm o}(t)\approx \omega_z\left (\frac{1}2+\frac{1}{\lambda} \right ).
\ee

Finally, before we proceed further, it is worth recalling here that the 
two lowest eigenstates of the global double-well potential can be well 
approximated as the symmetric and antisymmetric combinations of the local 
eigenstates i.e., 
$\ket{\pm}=(\ket{{\rm A}}\pm \ket{{\rm B}})/\sqrt{2}$, with  
eigenfrequencies $\omega_\pm=\omega_{\rm o}\pm J$. Hence, Hamiltonian 
\eref{Ht2} can be also expressed in terms of the corresponding symmetric 
and antisymmetric bosonic operators $\hat{d}_\pm=(\hat{a}\pm \hat{b})/\sqrt{2}$.
(e.g., see \cite{TwoModeValid}). Actually, in some cases one might get 
further detailed insights into the dynamics of the system if this is viewed in 
the basis of the global states.

\subsection{Outcoupled atoms}
Let us assume that atoms are coherently coupled out of the BEC A only. 
Neglecting collisions between trapped and free atoms, 
the many-body Hamiltonian for the free atoms is simply of the form \cite{jap00}
\be
\hat{\cal H}_{\rm f}=\int d{\bf r}\hat{\Psi}_{\rm f}^\dag({\bf r})
L_{\rm f}\hat{\Psi}_{\rm f}({\bf r}),
\ee
where 
\be
L_{\rm f}=-\frac{\hbar^2}{2m}\nabla +V_{\rm f}({\bf r}).
\ee

In general, the potential $V_{\rm f}({\bf r})$ experienced by the free atoms 
depends on the particular setup under consideration. Throughout this work 
we consider an atomic waveguide for the outcoupled atoms \cite{book,MHSguide}, 
resulting in an effective one-dimensional atom laser propagating 
along the weak confining axis of the waveguide 
(see figure \ref{system.fig}). For instance, 
such a guided atom laser has been demonstrated recently by 
Guerin {\em et al.} \cite{Gueprl06} and offers many advantages 
over the conventional outcoupling schemes. 
Formally speaking, the strong transverse confinement allows us to assume 
that the transverse dynamics of the free atoms adiabatically follow the 
slowly varying transverse potential of the optical guide 
$V_{\rm f}({\bf r}_\perp)$ \cite{Gueprl06}. 
For the sake of simplicity, throughout this 
work we assume that the transverse guide potential is nearly the 
same with the transverse potential of trap A i.e., 
$V_{\rm f}({\bf r}_\perp)\simeq V_{\rm A}^{\rm (L)}({\bf r}_\perp)$. 
In the absence of gravitational or other forces (as in the experimental 
setup \cite{Gueprl06}), the longitudinal component of the 
potential is $V_{\rm f}(z)=0$. 
Thus, the field operator for the free atoms can be expanded as 
\be
\hat{\Psi}_{\rm f}({\bf r}, t)=
\varphi_{\rm A}({\bf r}_\perp) \sum_k\chi_k(z)\hat{c}_k(t),
\label{Psi_f}
\ee 
where $\hat{c}_k$ is the annihilation operator of free atoms with momentum 
$\hbar k$ and obeys the usual bosonic commutation relations 
$\large [\hat{c}_k,\hat{c}_{k^\prime}^\dag \large ]=\delta_{k,k^\prime}$. 
The wavefunction $\varphi_A({\bf r}_\perp)$ is the ground-state 
wavefunction of the local transverse potential 
$V_{\rm A}^{\rm (L)}({\bf r}_\perp)$, with the normalization 
$\int d{\bf r}_\perp |\varphi_{\rm A}({\bf r}_\perp)|^2=1$, so that the 
linear atomic density $\rho_{\rm 1D}(z,t)\equiv
\int d{\bf r}_\perp|\hat{\Psi}_{\rm f}({\bf r}, t)|^2=
\sum_{k,q}\chi_k^*\chi_{q}\hat{c}_k^\dag\hat{c}_q$. 
The longitudinal wavefunction  $\chi_k(z)$ is readily obtained as a 
solution of the time-independent Schr\"odinger equation 
for a free atom (i.e., for $V_{\rm f}(z)=0$). Thus for a free atom 
with momentum $\hbar k$ we have $\chi_k(z)=e^{{\rm i}kz}/\sqrt{2\pi}$, 
and frequency 
\be
\omega_k=\frac{\hbar k^2}{2m}.
\label{disrel}
\ee
As we will see later on, this quadratic dependence of 
$\omega_k$ on $k$ is responsible for 
a number of mathematical difficulties arising in the context of atom 
lasers \cite{moypra299}. 
Using expansion (\ref{Psi_f}) and the orthonormality condition 
for $\chi_k(z)$, $\hat{\cal H}_{\rm f}$ reads 
\be
\hat{\cal H}_{\rm f}=\hbar\sum_k\left (\omega_k+
\frac{\omega_z}{\lambda}\right )\hat{c}_k^\dag \hat{c}_k. 
\ee

\subsection{Output coupling}
We consider an output coupling by application of external electromagnetic fields which induce an 
atomic transition from the internal state $(\ket{t})$ of the trapped atoms 
to an untrapped state $\ket{f}$. In the rotating-wave approximation, 
the many-body interaction Hamiltonian is of the form  \cite{jap00} 
\bea
\hat{\cal V}(t) &=& \hbar \int d{\bf r}
\hat{\Psi}_{\rm f}^\dag({\bf r})~\sqrt{\Lambda({\bf r},t)}~
\hat{\Psi}_{\rm t}({\bf r})+{\textrm{H.c}}
\label{H_couple_sw}
\eea
where $\Lambda({\bf r},t)$ is the coupling between trapped and untrapped atomic states. Using the expansions (\ref{Psi_t}) and (\ref{Psi_f}), we obtain
\bea
\fl
\hat{\cal V}(t) &=& \frac{\hbar}{\sqrt{2\pi}} \sum_k \hat{c}_k^\dag 
\int d{\rm r}\sqrt{\Lambda({\bf r},t)} 
\varphi_{\rm A}^*({\bf r}_\perp)e^{-{\rm i}k z}\left [ 
\varphi_{\rm A}({\bf r})\hat{a}(t)+\varphi_{\rm B}({\bf r})\hat{b}(t)
\right ]+
{\textrm{H.c}}
\eea        

In general the form of  $\Lambda({\bf r},t)$ depends on the particular 
outcoupling mechanism under consideration. 
Typical mechanisms may involve one-photon  radio-frequency  transition 
or indirect two-photon stimulated Raman transition 
\cite{bal96,adams,jap00,Graham,Edwards,MHSguide}. 
Note that in the following we neglect the momentum 
kick experienced 
by the atoms as well as the spatial dependence of 
$\Lambda({\bf r},t)$, obtaining 
\bea
\hat{\cal V}(t) =\hbar\sum_k g(k,t) \left (\hat{a}\hat{c}_k^\dag
+\hat{a}^\dag\hat{c}_k\right )+
\hbar e^{-\eta^2}\sum_k g(k,t) \left (
\hat{b}\hat{c}_k^\dag +\hat{b}^\dag\hat{c}_k\right ),
\label{Vh}
\eea        
where 
\be
g(k,t) = \frac{\sqrt{l_z}}{\pi^{1/4}}\sqrt{\Lambda(t)} e^{-k^2 l_z^2/2}.
\ee

According to equation (\ref{Vh}), the interaction consists of two terms  
despite the fact that the outcoupling mechanism is applied on BEC A only.
More precisely, the first term of $\hat{\cal V}(t)$ refers to BEC A 
and is similar to the expression used by many authors in the context of 
the standard 
single-mode model for the atom laser 
\cite{moypra299,hopepra97,moypra97,hopepra00,jackpra99,bre99,jefpra00,NLP03}. 
However, due to the presence of the second BEC, in our model we have obtained 
one more term which is proportional to the overlap of the 
two ground-state wavefunctions, that is $e^{-\eta^2}$.
In that respect, equation (\ref{Vh}) is a generalization of the single-mode 
outcoupling Hamiltonian \cite{moypra299,hopepra97,moypra97,hopepra00,jackpra99,bre99,jefpra00,NLP03}, to a two-mode scenario.

Let us now estimate the spectral response of the atomic continuum for the 
particular outcoupling mechanism under consideration. The density of 
states which are available to a free atom can be determined 
by the dispersion relation (\ref{disrel}) as follows
\be
\rho(\omega)=\bigg|\frac{dk}{d\omega}\bigg|=
	\sqrt{\frac{m}{2\hbar\omega}}\Theta(\omega),
\label{dos}
\ee
where $\Theta(\omega)$ is the usual step function. Note the divergence of the 
atomic density of states at the edge frequency $\omega_{\rm e}=0$, which is a
characteristic property of the one-dimensional model under consideration. 
Taking advantage of the 
symmetrical shape of the coupling and the even parity of $\omega_k$, 
we may reduce the $k$-space only to the $k>0$ sub-space. 
The spectral response of the continuum is then of the form
\be
D(\omega)=2|g(\omega,t)|^2\rho(\omega)=
\frac{\sqrt{2}\Lambda(t)}{\sqrt{\pi\omega_z}}
\frac{\exp(-2\omega/\omega_z)}{\sqrt{\omega}}\Theta(\omega),
\label{srG}
\ee
where $g(\omega,t)$ is readily obtained from $g(k,t)$ using the atomic 
dispersion relation (\ref{disrel}).

At this point we have completed the presentation of our model and the 
underlying approximations. 
In closing, let us summarize the main results by rewriting the 
complete form of the Hamiltonian under consideration 
in a frame rotating at $\omega_\perp$
\bea
\fl \hat{\cal H} = \frac{\hbar\omega_z}{2}(\hat{a}^\dag\hat{a}
+\hat{b}^\dag\hat{b})+\hbar\sum_k\omega_k\hat{c}_k^\dag \hat{c}_k
+\hbar \kappa (\hat{a}^\dag\hat{a}^\dag\hat{a}\hat{a}+
\hat{b}^\dag\hat{b}^\dag\hat{b}\hat{b})
+\hbar J(\hat{a}^\dag\hat{b}+\hat{b}^\dag\hat{a})\nonumber\\
\fl\quad\quad+\hbar\sum_k g(k,t) (\hat{a}\hat{c}_k^\dag
+\hat{a}^\dag\hat{c}_k )+
\hbar e^{-\eta^2}\sum_k g(k,t) (
\hat{b}\hat{c}_k^\dag +\hat{b}^\dag\hat{c}_k).
\label{fullham}
\eea

\section{Heisenberg equations of motion}
\label{SecIII}
Given the total Hamiltonian (\ref{fullham}) 
one may proceed to derive Heisenberg equations of motion 
for the operators of interest.
In the Heisenberg picture, the evolution of the expectation value 
of an arbitrary operator $\hat{\cal A}$ is governed by 
\[
\frac{{\rm d}\aver{\hat{\cal A}}}{{\rm d}t}=
-\frac{\rm i}{\hbar}\aver{[\hat{\cal A}, \hat{\cal H}]}.
\]
Thus,  for the operators pertaining to the two traps and the continuum, 
we obtain\\
\bea
\frac{{\rm d}\aver{\hat{a}}}{{\rm d}t}&=&-{\rm i}\frac{\omega_z}2\aver{\hat{a}}
-2{\rm i}\kappa\aver{\hat{a}^\dag\hat{a}\hat{a}}-{\rm i}J\aver{\hat{b}}
-2{\rm i}\int_0^\infty {\rm d}k g(k,t) \aver{\hat{c}_k},
\label{em1a}\\
\frac{{\rm d}\aver{\hat{b}}}{{\rm d}t}&=&-{\rm i}\frac{\omega_z}2\aver{\hat{b}}
-2{\rm i}\kappa\aver{\hat{b}^\dag\hat{b}\hat{b}}-{\rm i}J\aver{\hat{a}}
-2{\rm i}e^{-\eta^2}\int_0^\infty  {\rm d}k g(k,t) \aver{\hat{c}_k},
\label{em1b}\\
\frac{{\rm d}\aver{\hat{c}_k}}{{\rm d}t}&=&-{\rm i}\omega_k \aver{\hat{c}_k}
-{\rm i} g(k,t)\aver{\hat{a}}
-{\rm i} g(k,t)e^{-\eta^2}\aver{\hat{b}}. \label{em1c} 
\eea
We may now distinguish between two cases.

In the absence of interatomic interactions (i.e., for $\kappa=0$) the 
Hamiltonian (\ref{fullham}) becomes bilinear. As a result, the above set 
of equations is closed and all of the initial statistical 
properties of the system are preserved in time. For instance, 
if the BECs are initially in coherent states, we 
have 
$\aver{\hat{a}^\dag(t)\hat{a}(t)}=\aver{\hat{a}^\dag(t)}\aver{\hat{a}(t)}$, 
$\aver{\hat{b}^\dag(t)\hat{b}(t)}=\aver{\hat{b}^\dag(t)}\aver{\hat{b}(t)}$ 
and 
$\aver{\hat{c}_i^\dag(t)\hat{c}_j(t)}=\aver{\hat{c}_i^\dag(t)}\aver{\hat{c}_j(t)}$, 
for all $t\geq 0$ (see also \cite{hopepra97,moypra97,hopepra00,NLP03}). 
In other words, the  bilinear form of the Hamiltonian preserves 
the initial coherence in time, so that at any instant $t$ we can 
decorrelate exactly any higher-order correlation function  
in terms of $\aver{\hat{a}},\aver{\hat{b}}$, and $\aver{\hat{c}_k}$. 

In the presence of interatomic interactions 
(i.e., for $\kappa\neq 0$) the Hamiltonian (\ref{fullham}) 
involves fourth-order terms, and thus we have the appearance of third-order 
correlation functions in the right-hand side of equations 
(\ref{em1a})-(\ref{em1c}). This set of equations is no longer closed, 
while consideration of differential equations for the third-order 
correlation functions leads to the appearance of terms of even 
higher order and so on. In general, there are no exact remedies 
for such mathematical problems, but an approximate solution can be 
always obtained by decorrelating higher-order correlation functions into 
products of lower ones.
 
In the present work we decorrelate the third-order correlation 
functions appearing on the right-hand side of equations 
(\ref{em1a})-(\ref{em1c}) as follows:
$\aver{\hat{a}^\dag\hat{a}\hat{a}}\approx
\aver{\hat{a}^\dag}\aver{\hat{a}}\aver{\hat{a}}$  
and $\aver{\hat{b}^\dag\hat{b}\hat{b}}\approx 
\aver{\hat{b}^\dag}\aver{\hat{b}}\aver{\hat{b}}$. Hence, 
equations (\ref{em1a})-(\ref{em1c}) read 
\bea
\frac{{\rm d}\aver{\hat{a}}}{{\rm d}t}&=&-{\rm i}\frac{\omega_z}2\aver{\hat{a}}
-2{\rm i}\kappa |\aver{\hat{a}}|^2\aver{\hat{a}}-{\rm i}J\aver{\hat{b}}
-2{\rm i}\int_0^\infty  {\rm d}k g(k,t) \aver{\hat{c}_k},
\label{em1aD}\\
\frac{{\rm d}\aver{\hat{b}}}{{\rm d}t}&=&-{\rm i}\frac{\omega_z}2\aver{\hat{b}}
-2{\rm i}\kappa |\aver{\hat{b}}|^2\aver{\hat{b}}-{\rm i}J\aver{\hat{a}}
-2{\rm i}e^{-\eta^2}\int_0^\infty  {\rm d}k g(k,t) \aver{\hat{c}_k},
\label{em1bD}\\
\frac{{\rm d}\aver{\hat{c}_k}}{{\rm d}t}&=&-{\rm i}\omega_k \aver{\hat{c}_k}
-{\rm i} g(k,t)\aver{\hat{a}}
-{\rm i} g(k,t)e^{-\eta^2}\aver{\hat{b}}. \label{em1cD} 
\eea

One way to solve such a set of coupled differential equations 
is by means of the Laplace transform method. 
To this end, however,  one has to be able to perform all of the integrations 
over the continuum as well as the inverse Laplace transforms at the end. 
Both of these tasks are more or less straightforward in the case of smooth 
continua for which the Born and Markov approximations are applicable. 
In the present context, however, none of the aforementioned approximations 
is valid. Indeed, as we discussed earlier, the quadratic atomic dispersion 
relation is associated with a density of atomic states which diverges 
for small frequencies (see equation \ref{dos}). 
This behavior is also reflected in the spectral response (\ref{srG}) 
and implies that the continuum under consideration does not vary 
slowly for all frequencies. 

Structured continua which invalidate both Born and Markov approximations 
emerge in different areas of physics and have attracted considerable 
interest over the last few years \cite{bookBP}. 
To address fundamental mathematical difficulties associated 
with these continua a number of new theoretical techniques have been 
developed \cite{bookBP,bre99,str99,gara97,jack01}. 
Here, to deal with the structured continuum at hand, we follow a  
discretization approach developed in the context of photonic band-gap 
continua \cite{nikolg}. 
Briefly, we substitute the continuum for frequencies within a range 
around $\omega_{\rm o}$ (i.e., for $0<\omega<\omega_{\rm up}$), by a number 
(say $M$) of discrete modes, 
while the rest of the atom-mode density is treated perturbatively since 
it is far from resonance. Discussion on the choice of $\omega_{\rm up}$ 
and the number of discrete modes can be found in Refs. \cite{nikolg}.
This approach has been also applied in the 
context of atom lasers \cite{NLP03}, and is capable of providing 
not only the evolution of the number of atoms in the condensates, 
but also the distribution of the outcoupled atoms in frequency domain, 
irrespective of the strength of the outcoupling and the form of the 
spectral response. 

In general, a continuum can be discretized in many different ways 
(see for instance \cite{NLP03}) and 
in the present work we have chosen a uniform discretization scheme. 
In particular, we choose the frequencies of the discrete modes to be 
$\omega_j=j\varepsilon$, where the mode spacing $\varepsilon$ 
is determined by the upper-limit condition of the discretization, namely
$\omega_{\rm up}=M\varepsilon$. The corresponding coupling for the $j$ mode, 
is determined by the spectral response (\ref{srG}) as follows
\be
\tilde{g}_j^2=D(\omega_j)\varepsilon.
\label{disc1c}
\ee
Hence, working similarly to \cite{NLP03}, equations 
(\ref{em1aD})-(\ref{em1cD} ) read after the discretization
\bea
\fl \frac{{\rm d}\aver{\hat{a}}}{{\rm d}t}&=&
-{\rm i}\left (\frac{\omega_z}2-S\right )\aver{\hat{a}}
-2{\rm i}\kappa |\aver{\hat{a}}|^2\aver{\hat{a}}
-{\rm i}(J-Se^{-\eta^2})\aver{\hat{b}}
-{\rm i}\sum_{j=1}^{M} \tilde{g}_j \aver{\hat{c}_j},
\label{em1a2}\\
\fl \frac{{\rm d}\aver{\hat{b}}}{{\rm d}t}&=&
-{\rm i}\left (\frac{\omega_z}2-S e^{-2\eta^2}\right )\aver{\hat{b}}
-2{\rm i}\kappa |\aver{\hat{b}}|^2\aver{\hat{b}}
-{\rm i}(J-Se^{-\eta^2})\aver{\hat{a}}
-{\rm i}e^{-\eta^2}\sum_{j=1}^{M} \tilde{g}_j\aver{\hat{c}_j},
\label{em1b2}\\
\fl \frac{{\rm d}\aver{\hat{c}_j}}{{\rm d}t}&=&
-{\rm i}\omega_j \aver{\hat{c}_j}-{\rm i}\tilde{g}_j\aver{\hat{a}}
-{\rm i} \tilde{g}_j e^{-\eta^2} \aver{\hat{b}}, \label{em1c2} \eea
where
\bea
S&=&\int_{\omega_{\rm up}}^\infty\frac{D(\omega)}{\omega}d\omega.
\eea

\section{Simulations}
\label{SecIV}
Throughout our simulations we have considered $^{23}$Na BECs with 
$m=3.818\times 10^{-26}$ Kgr and $a_{\rm tt}=2.75\times 10^{-9}$ m, 
which are formed 
independently in identical harmonic traps with longitudinal oscillation 
frequency $\omega_z=200~\textrm{sec}^{-1}$ and ratio $\lambda=0.4$. 
We assume that the BECs A and B 
are initially prepared in coherent states $\ket{\alpha}$ 
and $\ket{\beta}$, respectively.
Equations (\ref{em1a2})-(\ref{em1c2}) are thus solved with initial conditions 
\be
\fl \aver{\hat{a}(0)}=\alpha=\sqrt{N\tilde{\alpha}(0)},
\quad 
\aver{\hat{b}(0)}=\beta=\sqrt{N\tilde{\beta}(0)} e^{{\rm i}\phi(0)}, 
\quad 
\aver{\hat{c}_j(0)}=0,
\label{init}
\ee
where $\phi(0)$ is the initial relative phase between the two BECs. 
Accordingly, the initial number of condensed atoms in the traps A and B are   
given by $N_{\rm A}(0)=|\aver{\hat{a}(0)}|^2=N\tilde{\alpha}(0)$ and 
$N_{\rm B}(0)=|\aver{\hat{b}(0)}|^2=N\tilde{\beta}(0)$. 
At any time $t\geq 0$ we have $\tilde{\alpha}(t)+\tilde{\beta}(t)+\tilde{\gamma}(t)=1$ 
so that $N_{\rm A}(t)+N_{\rm B}(t)+N_{\rm C}(t)=N$, where $N_{\rm C}(t)$ and  
$\tilde{\gamma}(t)$, respectively are the population and amplitude of the continuum. 
Finally, for the sake of simplicity and without introducing any significant 
errors, the applied outcoupling pulse $\Lambda(t)$ is modeled as rectangular 
lasting from $t=0$ to $t=\tau$. 

Most of the work on the non-Markovian aspects of atom-laser outcoupling 
has been performed in the framework of an ideal gas (i.e., for $\kappa=0$)
\cite{moypra299,hopepra97,moypra97,hopepra00,jackpra99,bre99,jefpra00,NLP03}. 
Hence, for the sake of comparison, in this section we focus mainly on 
the analysis of results obtained by simulations in non-interacting systems.
The case of weakly-interacting gases ($\kappa\neq 0$) deserves a thorough 
investigation and 
as such will be discussed in detail elsewhere. At the end of this section, 
however, we briefly summarize some of the main features of the 
weakly-interacting systems we have found in our simulations. 

\subsection{Ideal bosonic gas}  
As we mentioned earlier, in the absence of interatomic interactions 
no decorrelation approximations are necessary for the derivation 
of a  closed set of equations for the expectation values of the 
operators $\aver{\hat{a}(t)}$, $\aver{\hat{b}(t)}$ and $\aver{\hat{c}_j(t)}$. 
The evolution of the system is obtained by propagating equations 
(\ref{em1a2})-(\ref{em1c2}) with $\kappa=0$. As both BECs are assumed 
initially prepared in coherent states, all of the statistical properties of 
the system at any time $t$ can be expressed exactly in terms of 
$\aver{\hat{a}(t)}$, $\aver{\hat{b}(t)}$ and $\aver{\hat{c}_j(t)}$.

\subsubsection{Weak outcoupling---Markovian dynamics.} 
\begin{figure}
  \begin{center}
    \leavevmode
    \epsfxsize8.5cm
    \epsfbox{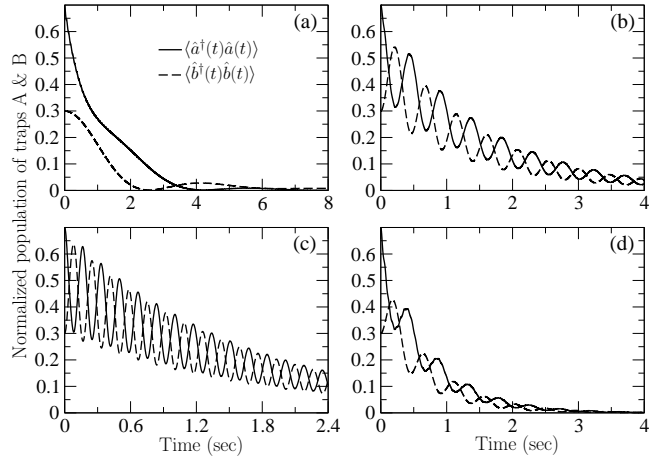}
  \end{center}
  \caption{Markovian regime. 
    Evolution of the normalized trap populations as a function of time 
    for various inter-trap distances and outcoupling strengths: 
    (a) $\Lambda=10^2~{\rm sec}^{-2}$, $\eta=2.0$;
    (b) $\Lambda=10^2~{\rm sec}^{-2}$, $\eta=1.7$;  
    (c) $\Lambda=10^2~{\rm sec}^{-2}$, $\eta=1.5$; 
    (d) $\Lambda=2\times 10^2~{\rm sec}^{-2}$, $\eta=1.7$. 
    System parameters: $\omega_z=200~{\rm sec}^{-1}$, $\lambda=0.4$.
    Initial conditions: $\tilde{\alpha}(0)=0.7$,  $\tilde{\beta}(0)=0.3$. 
    Discretization 
    parameters: $M=1500$, $\omega_{\rm up}=300~{\rm sec}^{-1}$.}
\label{markP:fig}
\end{figure}

As depicted in figure \ref{markP:fig}, for weak outcoupling strengths 
(i.e., for $\Lambda<5\times 10^2~\textrm{sec}^{-2})$, 
the dynamics of the system are mainly Markovian. More precisely, we have 
population exchange between the two traps, but the oscillations are 
exponentially damped as atoms are irreversibly coupled out of 
the traps. As we reduce the distance  $\eta$ between the traps, 
the Josephson coupling $J$ increases and the oscillations become faster 
(e.g., compare figures \ref{markP:fig}(b) and \ref{markP:fig}(c)). 
On the contrary, for constant intertrap distance, the oscillations decay faster 
as we increase the outcoupling rate $\Lambda$
(compare figures  \ref{markP:fig}(b) and \ref{markP:fig}(d)). 

\begin{figure}
  \begin{center}
    \leavevmode
    \epsfxsize8.5cm
    \epsfbox{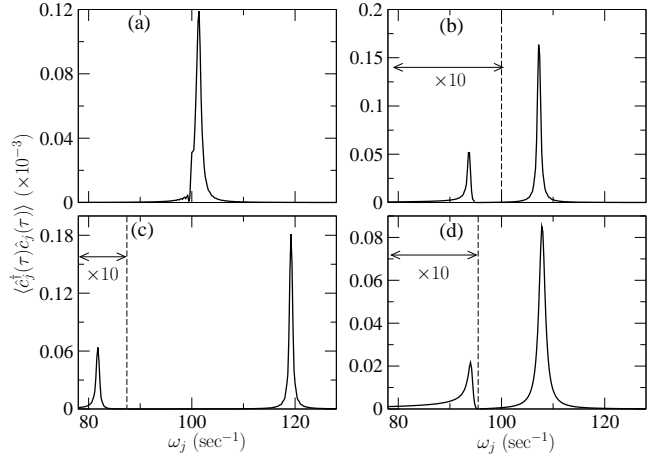}
  \end{center}
  \caption{Markovian regime. 
    Distribution of the outcoupled atoms at $\tau=10~{\rm sec}$ 
    for various inter-trap distances and outcoupling strengths. 
    The distributions (a)-(d) are in one-to-one correspondence with 
    figures \ref{markP:fig}(a)-(d), 
    respectively.
    Note that for the sake of illustration 
    the regime on the left-hand side of the vertical dashed line has been 
    magnified ten times.}
\label{markS:fig}
\end{figure}

In general, the distribution of the outcoupled atoms 
(see figure \ref{markS:fig}) exhibits the characteristic 
doublet which, however, is expected to be asymmetric mainly due to 
the unconventional density of atomic states (\ref{dos}). 
In particular, the origin of the doublet is well-described 
in terms of the global states $\ket{\pm}$ with eigenfrequencies 
$\omega_z/2\pm J$. 
The intertrap coupling splits the previously degenerate local states 
$\ket{{\rm A}({\rm B})}$ into a doublet of global states and thus the 
outcoupled atoms emerge as distinct peaks separated by $2J$. 
The outcoupling rate is larger for the global state with frequency closer 
to the edge $\omega_{\rm e}=0$, as the density of availlable atomic states, 
and thus the spectral response, scales as $1/\sqrt{\omega}$.  
Decreasing the distance between the two traps, 
the Josephson coupling increases monotonically for $\eta\leq 2.0$,  
and thus the distance between the peaks also increases 
(compare figures \ref{markS:fig}(b) and \ref{markS:fig}(c)). 
On the other hand, as depicted in the inset of figure \ref{system.fig}, 
for $\eta\geq 2.0$ the Josephson coupling is not strong enough to produce 
a noticeable splitting and to give rise to a clear doublet in the distribution 
of outcoupled atoms (see figure \ref{markS:fig}(a)). Finally, 
as we increase the outcoupling rate $\Lambda$ for constant $\eta$, 
the peaks become broader while their position remains practically unchanged 
(compare figures \ref{markS:fig}(b) and \ref{markS:fig}(d)) .

\subsubsection{Strong outcoupling---Non-markovian dynamics.} 
In the strong-outcoupling regime (i.e., for $\Lambda\geq 5\times 
10^2~\textrm{sec}^{-2}$) the trapped populations begin exhibiting 
non-Markovian dynamics. In general, the evolution of the system 
is governed by two different processes namely, the exchange of population 
between the BECs and the exchange of population between the BECs 
and the continuum. Note that exchange of population between any discrete 
feature and a continuum is a signature of the non-Markovian nature of 
the problem under consideration.  

\begin{figure}
  \begin{center}
    \leavevmode
    \epsfxsize8.5cm
    \epsfbox{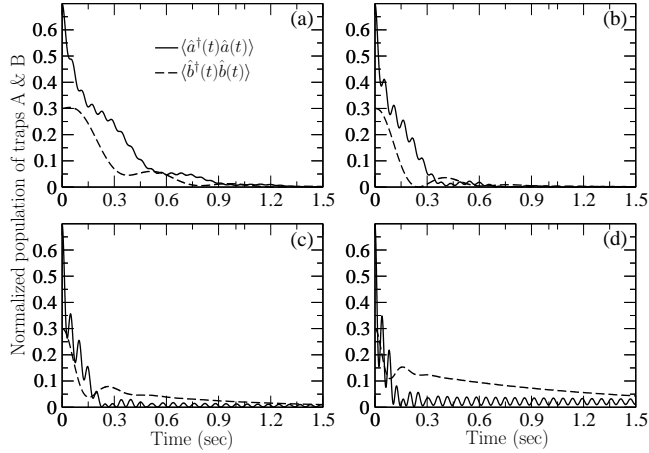}
  \end{center}
  \caption{Non-markovian regime. 
    Evolution of the normalized trap populations as a function of time 
    for various outcoupling strengths: 
    (a) $\Lambda=5\times 10^2~{\rm sec}^{-2}$;
    (b) $\Lambda=10^3~{\rm sec}^{-2}$;  
    (c) $\Lambda=2\times 10^3~{\rm sec}^{-2}$; 
    (d) $\Lambda=4\times 10^3~{\rm sec}^{-2}$. 
    System parameters: $\omega_z=200~{\rm sec}^{-1}$, $\lambda=0.4$, $\eta=1.7$.
    Initial conditions: $\tilde{\alpha}(0)=0.7$,  $\tilde{\beta}(0)=0.3$. Discretization 
    parameters: $M=1500$, $\omega_{\rm up}=300~{\rm sec}^{-1}$.}
\label{NMgam_pops:fig}
\end{figure}

In figure \ref{NMgam_pops:fig} we present the evolution of the trapped 
populations as functions of time, for a given intertrap distance 
and increasing outcoupling rate. Clearly, as far as trap B is concerned, 
we can identify an initial transient regime where the main part of the 
population is lost. After this initial stage, dissipation is temporarily turned 
off and trap B gets atoms from trap A (slight oscillations). This weak 
oscillatory population exchange between the two traps persists even for 
larger times,  but the population of trap B is gradually transferred into the 
continuum in an {\em irreversible}, almost exponential, way. 
Thus, irrespective of the 
strength of the outcoupling rate, trap B is always empty in the long-time limit 
(e.g., see figures \ref{NMgam_pops:fig}(a,b)). 
 
\begin{figure}
  \begin{center}
    \leavevmode
    \epsfxsize8.5cm
    \epsfbox{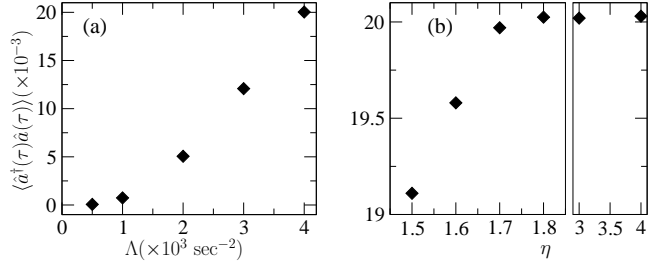}
  \end{center}
  \caption{Non-markovian regime. 
    Typical behavior of the steady-state population of trap A for varying  
    outcoupling rate (a) and dimensionless intertrap distance (b). 
    The depicted values are estimated numerically for a sufficiently dense 
    discretization and pulse duration $\tau=40$ sec. In the neighborhood of 
    $\eta\approx 2.14$, we have not found a steady-state for trap A, although 
    we have let the system evolve for times up to $\tau=50$ sec.
    System parameters: $\omega_z=200~{\rm sec}^{-1}$, $\lambda=0.4$, 
    $\eta=1.7$ for plot (a) and $\Lambda=4\times 10^3~{\rm sec}^{-2}$ 
    for plot (b).
    Initial conditions: $\tilde{\alpha}(0)=0.7$,  $\tilde{\beta}(0)=0.3$. 
    Discretization parameters: $M=3000$, $\omega_{\rm up}=300~{\rm sec}^{-1}$.}
\label{NMpopSS:fig}
\end{figure}

Moreover, in figure  \ref{NMgam_pops:fig} we see that, 
besides the weak oscillations, the population of trap A exhibits fast 
oscillations which become more pronounced and faster with increasing outcoupling 
strength. These oscillations are reflected only  in the population of the continuum 
which is not shown here. For relatively weak outcoupling rates not only trap B, 
but also trap A is empty in the long-time limit 
(e.g., see figures \ref{NMgam_pops:fig}(a,b)). On the contrary, for stronger  
outcoupling rates the system reaches a steady state pertaining to a practically 
empty trap B and a partly depleted BEC A 
(e.g., see figures \ref{NMgam_pops:fig}(d)). The formation of such a {\em 
bound state} 
has been demonstrated experimentally \cite{rmhc05} and involves atoms in a 
{\em superposition} of two states namely, the trapped and the untrapped state. 
The Born and Markov approximations are valid only if such superpositions 
decay on a time scale much shorter than the time scale of interest in 
this work. 
It is also worth noting that, according to our simulations 
(e.g., see figure \ref{NMgam_pops:fig}), the bound state involves 
only trap A and not trap B.  This is perhaps due to the weak outcoupling rate 
for BEC B which is 
proportional to the overlap $e^{-\eta^2}$ between the two BEC wavefunctions. 
Hence, BEC B is only weakly coupled to the continuum and any non-Markovian 
effects, such as the formation of a bound state, are suppressed.
Finally, although limitations in the 
validity of the two-mode model does not allow us to consider 
larger values of $\Lambda$, from figure \ref{NMpopSS:fig}(a) it is obvious 
that the population trapping increases as we increase the outcoupling rate. 
 
\begin{figure}
  \begin{center}
    \leavevmode
    \epsfxsize8.5cm
    \epsfbox{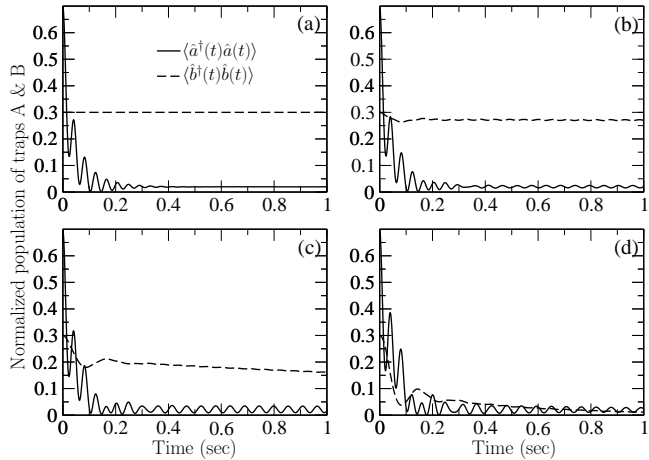}
  \end{center}
  \caption{Non-markovian regime. 
    Evolution of the normalized trap populations as a function of time  
    for various inter-trap distances:
    (a) $\eta=4.0$;
    (b) $\eta=2.0$;  
    (c) $\eta=1.8$; 
    (d) $\eta=1.6$. 
    System parameters: $\omega_z=200~{\rm sec}^{-1}$, $\lambda=0.4$, 
    $\Lambda=4\times 10^3~{\rm sec}^{-2}$.
    Initial conditions: $\tilde{\alpha}(0)=0.7$,  $\tilde{\beta}(0)=0.3$. Discretization 
    parameters: $M=1500$, $\omega_{\rm up}=300~{\rm sec}^{-1}$.}
\label{NMJ_pops:fig}
\end{figure}

Let us discuss now the effect of the intertrap distance $\eta$ 
on the evolution of the trapped populations. As depicted in figure 
\ref{NMJ_pops:fig} (a), for large intertrap distances the weak 
oscillations are absent and the population of trap B remains practically 
constant as the overlap between the BEC wavefunctions is negligible. 
As we bring the traps closer (see figures \ref{NMJ_pops:fig} (b-d)), 
the population of trap B  starts evolving in time 
with an initial transient regime followed by an irreversible decay. 
On the other hand, for the typical values of $\eta$ allowed by our two-mode 
approximation, there seems to be no significant effect on  
the evolution of the population of trap A. In particular, we have a continuous 
exchange of population between trap A and the continuum until the formation of 
a bound state in the long-time limit. Although the steady-state population of 
trap A does not change considerably over the regime of intertrap distances 
$\eta$ we can cover, a slight reduction is noticeable 
in figure \ref{NMpopSS:fig}(b) as we reduce $\eta$. 
Moreover, there is a regime of distances around $\eta_{\rm c} \approx 2.14$, 
where the system seems to have no steady state, in the conventional sense of 
the word, as it is beating between the two localized condensate modes 
even for the long time scale used 
in figure \ref{NMpopSS:fig}. That is why we do not give any values for 
$\aver{\hat{a}^\dag(\tau)\hat{a}(\tau)}$ in the neighborhood of 
$\eta_{\rm c}$.
All of this behavior can be easily understood if the system is viewed in the 
basis 
of the symmetric and antisymmetric global states $\ket{\pm}$. 

First of all let us briefly discuss the behavior of the global states 
$\ket{\pm}$ 
with eigenenergies $\omega_\pm=\omega_z/2\pm J$, as we approach the two traps. 
As depicted in the inset of figure \ref{system.fig}, the Josephson coupling  
and thus the separation of the global states, does not vary monotonically with 
$\eta$. More precisely,  as we reduce $\eta$ the splitting of the states 
$\ket{\pm}$ increases for $\eta \geq \eta_{\rm min}$, where 
$\eta_{\rm min}\approx 2.3$. 
In this regime of intertrap distances, $J$ is negative and thus 
$\omega_->\omega_+$ 
i.e., the symmetric state moves to the left and the antisymmetric state to the 
right of the characteristic frequency $\omega_z/2$. 
This relative movement is inverted for  $\eta_{\rm c}< \eta < \eta_{\rm min}$ and the 
two global states begin approaching each other. They become basically resonant 
at $\eta=\eta_{\rm c}$ where $J$ changes sign. Subsequently, as we further 
reduce 
$\eta$, the splitting of the states $\ket{\pm}$ increases again, 
but this time the global states move in opposite directions as $J>0$ and 
$\omega_+>\omega_-$. 
In particular, the symmetric state moves away from the edge and the population 
trapping associated with it decreases, whereas the antisymmetric state moves 
towards the edge where it is more protected against dissipation. In any case, 
it is worth noting that the Josephson coupling attains significant values 
only for $\eta<2.0$. Thus, for $2.0<\eta<3.0$ the two condensate modes are 
practically resonant, as the splitting is negligible. Hence, the system is 
beating between the two condensate modes and no steady-state is found on the 
time scale of figure \ref{NMpopSS:fig}.

In view of the above discussion, it is also easy to understand the reduction of the 
steady-state population of trap A as we bring the traps closer. 
For the particular initial conditions under consideration, we obtain 
$\aver{d_+^\dag(0)d_+(0)}=|\alpha+\beta|^2/2$ and 
$\aver{d_-^\dag(0)d_-(0)}=|\alpha-\beta|^2/2$. 
Throughout this work we have focused on in-phase BECs only i.e., $\phi=0$. 
As a result, the main part of the population initially occupies the 
symmetric global state and thus the behavior of the corresponding   
steady-state population with decreasing $\eta$ also determines the 
behavior of the steady-state population in the local state $\ket{\rm A}$.
This may not be the case if we choose different initial conditions, 
but the effect of the phase difference $\phi$ on the system's dynamics will 
be investigated in detail elsewhere.

\begin{figure}
  \begin{center}
    \leavevmode
    \epsfxsize8.5cm
    \epsfbox{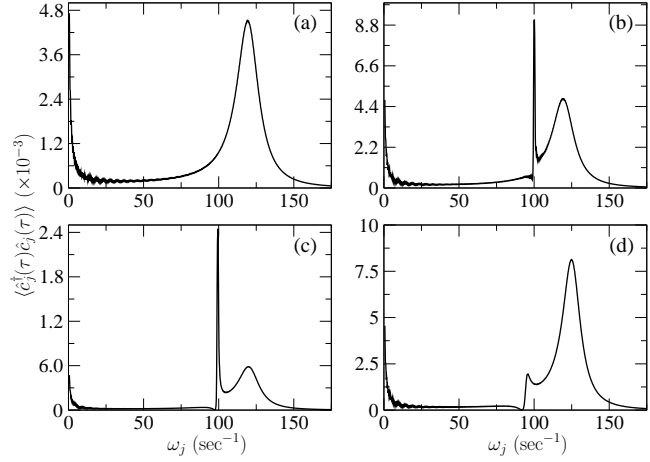}
  \end{center}
  \caption{Non-markovian regime. 
    Distribution of the outcoupled atoms at $\tau=10~{\rm sec}$ 
    for $\Lambda=2\times 10^3~{\rm sec}^{-2}$ and various 
    inter-trap distances: 
    (a) $\eta=4.0$;
    (b) $\eta=2.0$;  
    (c) $\eta=1.8$; 
    (d) $\eta=1.6$. 
    System parameters: $\omega_z=200~{\rm sec}^{-1}$, $\lambda=0.4$, 
    $\Lambda=2\times 10^3~{\rm sec}^{-2}$.
    Initial conditions: $\tilde{\alpha}(0)=0.7$,  $\tilde{\beta}(0)=0.3$. Discretization 
    parameters: $M=1500$, $\omega_{\rm up}=300~{\rm sec}^{-1}$.}
\label{NMJ_specs:fig}
\end{figure}

\begin{figure}
  \begin{center}
    \leavevmode
    \epsfxsize8.5cm
    \epsfbox{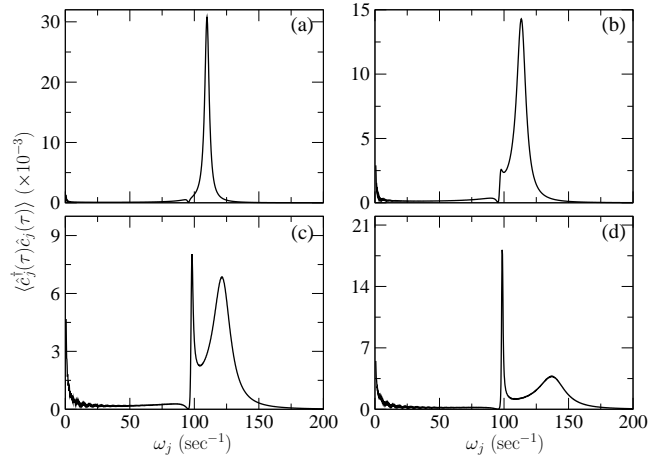}
  \end{center}
  \caption{Non-markovian regime. 
    As in figure \ref{NMJ_specs:fig} for $\eta=1.7$ and various outcoupling 
    strengths: 
    (a) $\Lambda=5\times 10^2~{\rm sec}^{-2}$;
    (b) $\Lambda=10^3~{\rm sec}^{-2}$;  
    (c) $\Lambda=2\times 10^3~{\rm sec}^{-2}$; 
    (d) $\Lambda=4\times 10^3~{\rm sec}^{-2}$. 
  }
\label{NMgam_specs:fig}
\end{figure}

We turn now to the discussion of the distribution of the outcoupled atoms 
in the non-Markovian regime.  
Unfortunately, for the spectral response (\ref{srG}) the derivation 
of analytic expressions for the atomic distribution is a rather difficult 
task \cite{moypra97}. In fact, analytic results can be obtained 
only in some special cases e.g., in the limit of broad-band output coupling,
by means of the Laplace transform method \cite{moypra97}. 
The discretization approach, however, is capable of providing us with the 
distribution of the outcoupled atoms at any time. 
In figures \ref{NMJ_specs:fig}  and \ref{NMgam_specs:fig}
we present such typical distributions at the end of the outcoupling pulse, 
i.e. at times $t=\tau$, with the pulse duration chosen sufficiently large 
to ensure that the distributions do not vary significantly with time.

For a better interpretation of these results, it is worth keeping in mind 
that the system 
under consideration involves two condensate modes which decay into the 
same atomic continuum. Moreover, there are two different 
outcoupling channels for each mode. More precisely, atoms can be 
coupled out of BEC A either directly or via BEC B and vice-versa; albeit 
at different rates. 
As a result we expect quantum interference effects which in addition to 
the non-Markovian nature of the dynamics may give rise to unconventional 
distributions of the outcoupled atoms. 

When the traps are far apart (see figure \ref{NMJ_specs:fig}(a)), 
we are essentially in the single-trap case 
where atoms are coherently outcoupled from BEC A only. 
In this case any quantum interference effects are absent and the distribution 
of the outcoupled atoms  exhibits a  
well-known profile previously discussed in Ref. \cite{NLP03}. 
More precisely, we have a peak around the condensate-mode frequency 
$\omega_z/2$, and a peak around the edge frequency $\omega_{\rm e}=0$  
where the density of atomic states diverges (see equation \ref{dos}). 
The former peak is shifted towards higher frequencies due to the 
coupling between the condensate mode and the continuum. 
This effect has also been noted by other authors  
(e.g., see \cite{jefpra00}) and becomes more pronounced as we increase the 
outcoupling rate. Moreover, the time-dependence of the presented distributions 
gives rise to oscillations which become faster as we increase the pulse 
duration. In the Markovian regime, the atomic distributions do not 
exhibit any such oscillations (see figure \ref{markS:fig}) as all of the 
time-dependent terms entering 
the distributions become practically negligible for times $\Lambda t\gg 1$. 
In the non-Markovian regime, however, we have a continuous exchange of atoms 
between the BECs and the continuum even for larger times as well as the 
formation of a non-decaying bound mode in the long-time limit. 
Hence, there exist time-dependent terms associated with the formed bound mode 
which persist even in the limit $t\to\infty$ and their effect on the atomic  
distribution is evident, unless one performs a time average over the period 
of the oscillations (see also discussion in \cite{bay97,ebewod77}). 

As the traps approach each other (see figure \ref{NMJ_specs:fig}(b)), 
the overlap between the two BECs increases and a narrow peak appears next 
to the main peak. This narrow peak is clearly associated with a rather weak 
outcoupling and thus can be attributed to atoms originated from trap B. 
It could be said therefore that for intermediate intertrap distances 
the distribution of the outcoupled atoms is basically a 
superposition of the distributions for each individual trap. 
As we reduce further the intertrap distance $\eta$, quantum interference 
effects start becoming significant, and a clear dip appears next to the 
narrow peak (see figures \ref{NMJ_specs:fig}(c,d)). 

\begin{figure}
  \begin{center}
    \leavevmode
    \epsfxsize8.5cm
    \epsfbox{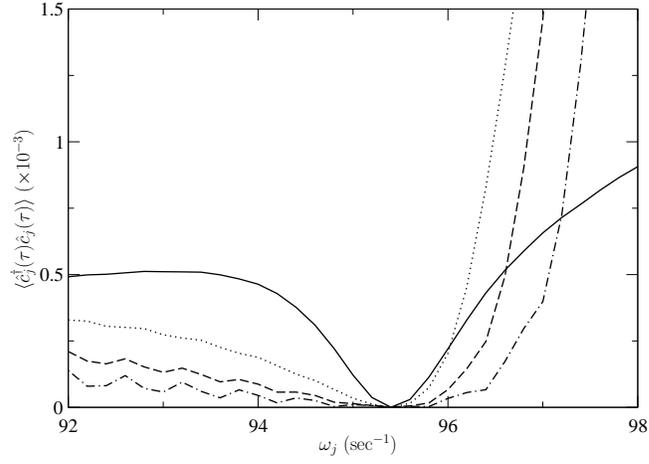}
  \end{center}
  \caption{Non-markovian regime. A closeup of the atomic distribution 
    around the dip for the parameters of figure \ref{NMgam_specs:fig} and various 
    outcoupling strengths: $\Lambda=5\times 10^2~{\rm sec}^{-2}$ (solid line), 
    $\Lambda=10^3~{\rm sec}^{-2}$ (dotted line),  
    $\Lambda=2\times 10^3~{\rm sec}^{-2}$ (dashed line), and  
    $\Lambda=4\times 10^3~{\rm sec}^{-2}$ (dot-dashed line). 
  }
\label{NMdip:fig}
\end{figure}

As depicted in figure \ref{NMgam_specs:fig}(a), the dip is also present 
for relatively weak outcoupling strengths provided the two traps are close 
enough. Actually, it was also present in the Markovian regime 
(see figure \ref{markS:fig}), but it was not so clear 
as the two peaks were far apart. It is obvious therefore that the dip is a 
clear evidence of destructive interference between the various outcoupling 
channels of the system. Although the outcoupling itself is not sufficient  
to cause the dip, in figure \ref{NMdip:fig} we see that the 
dip becomes broader with increasing outcoupling rates, while one may also 
notice the emergence of the narrow peak at frequencies around 
$\omega_z/2$ in figures \ref{NMgam_specs:fig}(b)-(d).

Due to the lack of analytic expressions, it is not clear whether the observed 
dip at $\omega_j\equiv\omega_{\rm d}$ is a perfect dark line i.e, 
whether $\aver{\hat{c}_j^\dag(\tau)\hat{c}_j(\tau)}=0$ at $\omega_{\rm d}$. 
To resolve this issue we have obtained analytic 
expressions for the distributions in the broad-band limit of the output 
coupling where 
$D(\omega)\sim \Lambda(t)/\sqrt{\omega}$ \cite{moypra299,moypra97}. 
In this limit, the observed profiles are similar to the ones presented here  
and we can verify that the atomic distribution at 
$\omega_j=\omega_{\rm d}$ vanishes only in the limit $t\to\infty$. 
Otherwise, for $\Lambda t\gg 1$ the dip corresponds to a very low probability 
($\sim 10^{-6}$) for outcoupled atoms with frequency $\omega_{\rm d}$. 

\subsection{Weakly-interacting bosonic gas}
Considering a harmonic trap with longitudinal oscillation frequency 
$\omega_z=200\textrm{sec}^{-1}$ and ratio $\lambda=0.4$, 
condition (\ref{Nmax}) yields $N\ll 2.5\times 10^{3}$. 
In other words, our model is valid for small BECs consisting of a 
few hundred of atoms. 
As we discussed in section \ref{SecIII}, in the presence of interactions 
we can obtain a closed set of equations of motion for the operators 
$\aver{\hat{a}(t)}$, $\aver{\hat{b}(t)}$ and $\aver{\hat{c}_j(t)}$, only by 
applying a decorrelation approximation. In view of this decorrelation, 
interatomic interactions enter the equations of motion for 
$\aver{\hat{a}(t)}$ and $\aver{\hat{b}(t)}$ as time-dependent shifts 
proportional to the corresponding trapped populations 
$N_{\rm A}(t)=|\aver{\hat{a}(t)}|^2$ and $N_{\rm B}=|\aver{\hat{b}(t)}|^2$ 
(see equations \ref{em1a2} and \ref{em1b2}). 
As a result, the frequencies of the condensate modes A and B fluctuate 
in time and become off-resonant. 

The evolution of the weakly interacting system is governed by 
three distinct physical processes. More precisely, apart from the  
Josephson and the output coupling which were also present in the 
interaction-free model, we also have the repulsive collisional 
interactions. It is reasonable therefore to define the ratios 
$N_{\rm c}^{({\rm t})}=J/\kappa$ and 
$N_{\rm c}^{({\rm f})}=\sqrt{\Lambda}/\kappa$ 
which quantify the effect of interatomic interactions relative
to tunneling and outcoupling effects, respectively. 
How strongly the inclusion of interactions affects the 
results obtained in the framework of the interaction-free model depends on 
the ratios $N/N_{\rm c}^{({\rm t})}$ and $N/N_{\rm c}^{({\rm f})}$, where 
$N$ is the total number of atoms in the system. 

The case of weakly-interacting gases is of particular interest but it 
cannot be covered in the present work as there are many aspects which 
need to be thoroughly investigated. For instance, it is already known  
that for an isolated 
double-well BEC (i.e., in the absence of losses and outcoupling) one may 
define two extreme regimes of dynamics \cite{SelfTrapping} 
namely the Josephson regime \cite{java86} and the self-trapping regime 
\cite{SelfTrapping} . The detailed analysis of the previous section 
does not involve self-trapping at all as this phenomenon occurs only 
for Hamiltonians involving interactions. 
Here, we would like to briefly highlight only some of the features of the 
weakly-interacting model we have found in our simulations. 
The detailed presentation and discussion of the results will be the 
subject of a forthcoming work. 

In the strong outcoupling regime (i.e., for $N_{\rm c}^{({\rm f})}>N$), 
inclusion of interactions affects the evolution of the trapped populations 
only quantitatively. The most important features of the 
weakly-interacting system in this regime seem to be the destruction of 
the bound 
mode and the disappearance of the dark spectral line discussed in the 
context of the interaction-free model. 
On the other hand, for $N_{\rm c}^{({\rm f})}<N$ interatomic interactions 
dominate over the output coupling and thus we have strong modifications 
in the evolution of the populations as well as the distribution of the 
outcoupled atoms.

\section{Summary and outlook}
\label{SecV}
We have investigated the non-Markovian aspects of atom-laser outcoupling 
from a 
double-well BEC. Our two-mode trapped condensate model has been motivated 
by recent 
experiments on the merging of independently formed BECs \cite{MITscience02}, 
and the first realization of a guided quasicontinuous atom laser 
\cite{Gueprl06}. 
For the sake of comparison with earlier work, relying on a single-mode trapped 
condensate, we have focused on an interaction-free model. 
In particular, we have studied how the presence of the second BEC (BEC B)  
affects the evolution of the trapped populations as well as the distribution 
of the outcoupled atoms. Although the outcoupling mechanism is 
basically applied only to BEC A, atoms are also weakly outcoupled from BEC B 
due to the overlap between the BEC wavefunctions.

In the case of weak outcoupling rates the dynamics of the system are 
purely Markovian and thus particularly simple. More precisely, the system 
oscillates between the two condensate modes while atoms are coherently 
outcoupled into the continuum. The distribution of the 
outcoupled atoms exhibits the characteristic asymmetric doublet due to 
the unconventinal density of atomic states and the different outcoupling 
rates experienced by the BEC atoms. 

The situation is substantially different in the strong-outcoupling regime, 
where the system exhibits non-Markovian dynamics. In particular, we have 
population exchange between the two BECs as well as between the BECs and the 
continuum. In the latter process, BEC B seems to participate passively as  
its population is gradually transferred to the continuum in an irreversible 
way. 
On the contrary, BEC A keeps exchanging population with the continuum even for 
larger times while for sufficiently strong outcoupling rates it is found only 
partially depleted in the long-time limit. The formation of such a bound state 
has also been predicted in the context of single-mode models. 
However, in the two-mode model under consideration the long-time-limit 
behavior 
of the system seems to depend on several parameters such as the 
outcoupling rate, 
the intertrap distance and the phase difference between the two BECs. 

The non-Markovian nature of the dynamics and the presence of the second BEC 
are mostly apparent in the distribution of the outcoupled atoms which exhibits 
two peaks. On the one hand, there is a broad peak stemming from atoms directly 
outcoupled from BEC A while on the other hand, atoms outcoupled from BEC B 
give rise to a rather narrow peak. Most importantly, as a result of destructive 
quantum interference between various outcoupling channels in the system, the 
atomic distribution may also exhibit a dark spectral line. It is worth noting 
that similar quantum interference phenomena have been discussed in the context 
of optical systems. Nevertheless, to the best of our knowledge the appearance of 
dark lines in the spectrum of atom lasers has not been discussed in the 
literature so far. 
 
In general, the two-mode model considered here allows for the direct 
inclusion of collisional interactions between trapped atoms. 
Nevertheless, due to space limitations, throughout this work we have focused 
on an interaction-free model only, and have assumed that the two BECs are 
initially in phase. 
Effects of interactions and the role of the phase difference 
will be presented in detail elsewhere.

\section*{Acknowledgments}

The authors thank David Petrosyan for his critical reading of the manuscript.

\Figures

\end{document}